\documentclass[twoside]{article}
\usepackage{fleqn,espcrc2}

\usepackage{epsfig}
\usepackage{graphicx}
\usepackage[figuresright]{rotating}

\newcommand{\be}{\begin{equation}}
\newcommand{\ee}{\end{equation}}
\newcommand{\ba}{\begin{eqnarray}}
\newcommand{\ea}{\end{eqnarray}}
\newcommand{\<}{\langle}
\renewcommand{\>}{\rangle}

\newcommand{\AmS}{{\protect\the\textfont2
  A\kern-.1667em\lower.5ex\hbox{M}\kern-.125emS}}
\def\spose#1{\hbox to 0pt{#1\hss}}
\def\ltapprox{\mathrel{\spose{\lower 3pt\hbox{$\mathchar"218$}}
 \raise 2.0pt\hbox{$\mathchar"13C$}}}
\def\ZP{{ Z.\ Phys.\ }}
\def\PR{{ Phys.\ Rev.\ }}

\def\PL{{ Phys.\ Lett.\ }}

\def\lsim{\raise0.3ex\hbox{$<$\kern-0.75em\raise-1.1ex\hbox{$\sim$}}}
\def\gsim{\raise0.3ex\hbox{$>$\kern-0.75em\raise-1.1ex\hbox{$\sim$}}}

\title{Short distance physics with heavy quark potentials
       \thanks{This work was supported by DFG Grant No. FOR 339/1-2.}}

\author{F. Zantow with O. Kaczmarek, F. Karsch, P. Petreczky \\\vspace{0.5cm}
       {Fakult\"at f\"ur Physik, Universit\"at Bielefeld, \\
        D-33615 Bielefeld, Germany}}

\begin{document}
\begin{abstract}
We present lattice studies of heavy quark potentials in the quenched
approximation of QCD at finite temperatures. Both, the color singlet and
color averaged potentials are calculated. While the potentials are well known
at large distances, we give a detailed analysis of their short distance
behavior (from 0.015 fm to 1 fm) near the critical temperature.
At these distances we expect that the T-dependent potentials go over into
the zero temperature potential. Indeed, we find evidences that the temperature
influence gets suppressed and the potentials starts
to become a unique function of the underlying distance scale. We use this feature to
normalize the heavy quark potentials at short distances and extract the free
energy of the quark system in a gluonic heat bath.

\end{abstract}
\maketitle
\section{INTRODUCTION}
The study of heavy quark potentials (free energies) is important for our understanding
of the fundamental forces in strongly interacting matter.
 For example it finds applications in the phenomenology of heavy
quarkonium physics at finite temperature [1,2,3]. Our aim is to get deeper
insights in their short distance behavior ($R<0.1fm$). It is
expected that at distances $R<<1/T$ the presence of a thermal medium does not
modify physics. However the question arises at which distance scales the temperature
effects become negligible.

\indent We performed lattice studies of the color averaged and the color
singlet heavy quark free energy. Our calculations have been performed on a $32^3\times 8$
and $32^3\times 16$ lattice using a tree-level Symanzik improved action and on
a $64^3\times 16$ lattice using the standard Wilson action of the pure $SU(3)$
theory. Up to an additive renormalization, the color averaged
free energy, $F_{av}:=F_{av}(R,T)$, is defined in terms of Polyakov loop
correlation functions [4] $F_{av}=-T\ln
\<\tilde{\mbox{Tr}}L(\vec{0})\tilde{\mbox{Tr}}L^\dagger(\vec{R})\> + C_{av}$,
where $L$ denotes the Polyakov loop and $\tilde{\mbox{Tr}}$ is the normalized
trace: $\tilde{\mbox{Tr}}\equiv 1/N_c\mbox{Tr}$. Following Ref.~[5] the color
singlet free energy $F_s:=F_s(R,T)$ is identified with the cyclic Wilson loop via
$F_s=-T\ln \<\tilde{\mbox{Tr}}L(\vec{0})D_{\vec{R}}L^\dagger(\vec{R})D_{\vec{R}}^\dagger\> + C_s$,
where $D_{\vec{R}}$ is the product of links on the shortest connection from
$\vec{0}$ to $\vec{R}$. This formula is expected to be valid at short distances
 [5]. The renormalization constants $C_{av}$ and $C_s$ contain temperature
 independent divergent self-energy contributions. As no additional divergences
 are introduced at $T>0$ we can fix the renormalization constants, if we are
 able to separate the $T$-independent part in the free energy and match it to
 the zero temperature heavy quark potential.

\indent The color averaged free energy can be related to the singlet ($F_s$) and
octet ($F_o$) contributions due to:
\begin{eqnarray}
&&\exp(-F_{av}/T)\nonumber\\
&&=1/9\;\exp(-F_s/T)+8/9\;\exp(-F_o/T).
\end{eqnarray}
In a static system the internal energy reduces to the potential $V$ and
thus the free energy refers to $F=V-TS$, where $S$ denotes the entropy.
In a perturbative treatment of QCD at lowest order the singlet and octet
potentials are of Coulomb form [6]:
\begin{eqnarray}
V_s(R)=-\frac{4}{3}\;\frac{g^2}{4\pi R}\,,\qquad V_o(R)=-\frac{1}{8}V_s(R),
\end{eqnarray}
where $g$ is the QCD running coupling. A high temperature expansion of (1) in
combination with (2) yields the color averaged heavy quark potential:
\begin{eqnarray}
\frac{V(R,T)}{T}\;=\;-\frac{1}{9}\;\frac{g^4}{(4\pi)^2}\frac{1}{(RT)^2}\;=\;
-\frac{\alpha_s^2}{9(RT)^2},
\end{eqnarray}
with $\alpha_s=g^2/(4\pi)$. Note that (3) is restricted to large
distances in order to justify the high temperature expansion. At short
distances the octet contribution in (1) is suppressed leading to a singlet-like
behavior of the color averaged potential [7].

\section{THE COLOR AVERAGED POTENTIAL AT $T>T_c$}
\label{section:results}
We study the short distance behavior of the color averaged potential in terms
of the screening function $S(R,T)$ defined by $S(R,T)=-9V(R,T)T[1/R]^{-2}$,
with $[\frac{1}{R}]$ being the lattice Coulomb potential. At large
distances the screening function is exponentially suppressed (hence its name)
and in the regime where the perturbative expression (3) is applicable we expect
$S=\alpha_s^2$. In Fig.~1 the square root of the screening function
is plotted. Two distinct features are evident from our data: First, the screening function is decreasing at very
short distances ($RT\lsim 0.1$)  indicating the
singlet-like behavior mentioned before. Secondly, screening effects become important at
rather short distances (about $0.3/T$). Nonetheless, there still remain temperature effects
in the color averaged potentials at short distances.

We have also compared the lattice data
with a 1-loop perturbative calculation (up to ${\cal O}(g^6)$) in Fig.~1. In this calculation we
have added the non-static contributions to the potential which are expected to become
important in the short distance regime (see Appendix A). As one can see from
Fig.~1 the color averaged
potential can be described by 1-loop perturbation theory for
$T\gsim 6T_c$ and distances $0.3\lsim RT\lsim 1$. For smaller temperatures and larger
distances perturbation theory breaks down due to non-perturbative effects.
Our calculation also breaks down at very small distances $RT\lsim 0.1$ because of the
enclosed high temperature expansion.

\begin{figure}[ht]
\begin{center}
   \epsfig{file=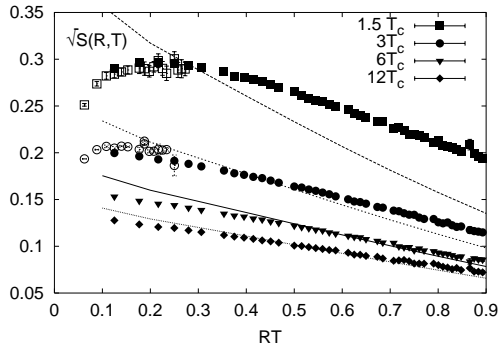,width=70mm}
\end{center}
\vspace{-1.5cm}
\caption{The square root of the screening function $\sqrt{S}$ calculated on a
  $32^3\times 8$ lattice (filled symbols) and on a $32^3\times 16$ (or
  $64^3\times 16$) lattice (open symbols) and the corresponding perturbative
  results (lines).}
\end{figure}

\section{THE TEMPERATURE INDEPENDENCE AT SHORT DISTANCES}
\label{section:results}
In order to eliminate effects from color averaging we now refer to the
short distance behavior of the color singlet potentials.
Our results together with the zero temperature
potential from [8] are shown in Fig~2. The potentials have been normalized at
short distances to make a comparison with the zero temperature limit possible.
As one can see from Fig.~2 the
temperature dependence of the heavy quark potential becomes first visible at
distances of about $R\gsim 0.1fm$. Thus the (finite $T$) color singlet potentials approach
their zero temperature limit already at the plotted distances. Indeed this
analysis provides evidence that the running coupling extracted from color singlet
data using (2) starts to get $T$-independent at these distances.

We now argue that these normalized data refer to the free energy of
the estimated quark system - enclosing effects from the gluonic heat bath - because divergent self-energy contributions cancel
out. Moreover we expect that the free energy has a well defined
continuum limit in our approach because the normalization gets
independent of the lattice cut-off at sufficiently short distances (see Fig.~2).

\begin{figure}[ht]
\begin{center}
   \epsfig{file=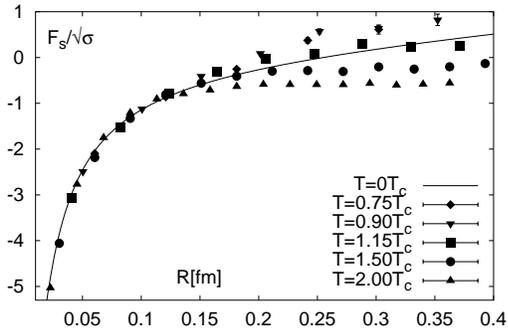,width=70mm}
\end{center}
\vspace{-1.5cm}
\caption{The zero temperature limit of color singlet potentials at non-zero temperature.}
\end{figure}

\section{CONCLUSIONS}
\label{section:results}
Our study shows that the heavy quark potentials have a non-trivial structure at
short distances. Color averaging influences the physics of potentials at short
distances and causes that the averaged data approach the $T=0$ limit at shorter
distances as do the singlet data.
The color singlet potentials approach the zero temperature limit at distances
$R\sim 0.1fm$ and $T\sim T_c$.
Due to this feature we have extracted the free energy of a heavy quark pair
enclosing the gluonic heat bath. A comparison
of our 1-loop perturbative calculation with lattice data suggests that the
perturbative expansion is applicable for temperatures $T\gsim 6T_c$ and distances
$0.3\lsim RT\lsim 1$.

\begin{appendix}
\section{THE NON-STATIC CONTRIBUTIONS}
\label{section:results}
We have calculated the color averaged heavy quark potential
at finite temperature based on $S[A_i,A_0]=\int dx_0\int d^3x
(-1/2(F^a_{0i})^2-1/4(F^a_{ij})^2)$. In this expression
$F^a_{\mu\nu}$ is the usual non-abelian field-strength tensor where $A_i$
($i=1,2,3$) denotes the static and $A_0$ the non-static gauge field and we
introduce a temporal gauge ($\partial_0A_0=0$).
The $g^6$-correction (1-loop) to the color averaged potential can be divided in
a static ($\Delta V^{st}$) and a
non-static ($\Delta V^{ns}$) part:
\begin{eqnarray}
 \Delta V_{av}(R,T)&=&\Delta
V^{st}(R,T)+\Delta V^{ns}(R,T).
\end{eqnarray}
The static contribution is well known from [5].
For the non-static part we find at 1-loop level:
\begin{eqnarray}
&&\Delta V^{ns}(R,T)=\nonumber\\
&&-\frac{g^4\beta^2(N^2-1)}{N^2}\left(\frac{e^{-m_DR}}{4\pi R}\right)\nonumber\\
&&\times \int\frac{d^3p}{(2\pi)^3}\left(\Pi^{ns}_{00}(p)-m_D^2\right)\frac{1}{
(p^2+m_D^2)^2}e^{ipR},\nonumber\\
\end{eqnarray}
where $\Pi^{ns}_{00}(p)$ is the self-energy of the non-static field with $\Pi^{ns}_{00}(p=0)=m_D^2$ and $\beta=1/T$. Note that in our conventions the
leading order term comes from the non-static contribution.
In finite temperature field theory $\Pi_{00}^{ns}(p)$ consists of the UV-divergent
zero temperature part and the convergent thermal part: $\Pi^{ns}_{00}(p)=\Pi^{(T=0)}_{00}(p)+\Pi^{(T)}_{00}(p)$.
$\Pi_{00}^{(T=0)}(p)$ is calculated in [9]:
\begin{eqnarray}
&&\Pi_{00}^{(T=0)}(p)\nonumber\\
&&=-\frac{g^2N}{16\pi^2}\left(\frac{11}{3}\ln\frac{4\pi\nu^2e^{-\gamma_E}}
{p^2}+\frac{31}{9}\right)p^2.
\end{eqnarray}
The thermal contribution can be extracted from a direct integration of the
loop diagrams contributing to $\Pi_{00}^{ns}(p)$ with standard calculation
techniques at finite temperature. Our 1-loop calculation of $\Pi_{00}^{ns}(p)$
is in accordance with (6) and the resulting potentials are plotted in Fig.~1.

\end{appendix}

\end{document}